\documentclass[preprint,showpacs,amsmath]{revtex4}
\usepackage[dvips]{graphicx}
\usepackage{comment}
\usepackage{amsmath}
\usepackage{xspace}
\usepackage{ulem}
\usepackage{subfigure}
\usepackage{placeins}
\usepackage{cancel}
\usepackage{color}
\newcommand{\flabel}[1]{\label{fig:#1}}

\begin{document}

\title{
Scaling in the Timing of Extreme Events 
}

\author{ \'Alvaro Corral$^{1,2}$}

\affiliation{
$^1$Centre de Recerca Matem\`atica, Campus de Bellaterra, Edifici C, E-08193 Barcelona, Spain\\
$^2$Departament de Matem\`atiques, Universitat Aut\`onoma de Barcelona, E-08193 Barcelona, Spain\\
}
\date{\today}
\begin{abstract}
%... OJO A LO QUE SE DICE DE SOC AQUI!!!!
Extreme events can come either from point processes, when the size or energy
of the events is above a certain threshold, or from time series, when the intensity of a signal
surpasses a threshold value. We are particularly concerned by the time between these extreme events, 
called respectively waiting time and quiet time. 
If the thresholds are high enough it is possible to justify the existence of scaling laws for
the probability distribution of the times as a function of the threshold value, although the scaling functions
are different in each case.
For point processes, in addition to the trivial Poisson process, one can obtain double-power-law distributions
with no finite mean value. This is justified in the context of renormalization-group transformations, 
where such distributions arise as limiting distributions after iterations of the transformation.
Clear connections with the generalized central limit theorem are established from here.
The non-existence of finite moments leads to a semi-parametric scaling law in terms of
the sample mean waiting time, in which the (usually unkown) scale parameter is eliminated
but not the exponents. 
In the case of time series, scaling can arise by considering random-walk-like signals with absorbing boundaries,
resulting in distributions with a power-law ``bulk'' and a faster decay for long times.
For large thresholds the moments of the quiet-time distribution show a power-law dependence 
with the scale parameter, and isolation of the latter and of the exponents leads to a non-parametric scaling
law in terms only of the moments of the distribution.  
Conclusions about the projections of changes in the occurrence of natural hazards
lead to the necessity of distinguishing the behavior of the mean of the distribution
with the behavior of the extreme events.
\end{abstract}

%\pacs{
%05.65.+b, %%Self-organized systems
%05.70.Jk, %%Critical point phenomena
%64.60.Ht %%Dynamic critical phenomena
%92.40.E- %%Precipitation (see also 92.60.jf?in meteorology)
%}

\maketitle

%%PASAR CORRECTOR!!!!!

\section{Introduction}

The theory of self-organized criticality (SOC), introduced by 
Bak and collaborators, provides an appealing explanation
for the distribution of sizes (and presumably durations)
of many naturally occurring catastrophic events \cite{Bak_book,Pruessner_book}.
Earthquakes, %forest fires, 
landslides, volcanic eruptions, 
%solar flares, 
rainfall, hurricanes,
etc. 
\cite{Bak_book,Malamud_hazards,Peters_Deluca,Corral_hurricanes}, 
have been found to 
display power-law distributions of event sizes,
which implies the absence of characteristic scales
for those sizes.
This means that if you ask, for instance:
{``how big are earthquakes in Japan?''},
this innocent question has no possible answer
\cite{Christensen_Moloney,Corral_Lacidogna}.
In mathematical terms, this implies that after
performing a scale transformation of a function
describing the size of the events,
the function has to remain unchanged,
and the only univariate function with this property
(for any dilatation or contraction)
is the power law.

SOC explains this power-law behavior 
as coming from a self-organization process
that leads the system to the critical point of
a phase transition, where scale invariance
is ensured.
Other approaches, purely probabilistic, 
support the power-law tail in some cases, 
as described by the generalized Pareto distribution \cite{Coles}.
Nevertheless, the exponent of the power law
in self-organized-critical systems
takes such values that it is associated to an infinite expected value
for the size of the events, which is unphysical; 
therefore, in practice, one cannot ignore finite-size effects.

When one is worried not about the size but about the time between events, 
the situation is a bit more involving.
In fact, the waiting time %, or the quiet time, 
is defined as the time between consecutive events
when these events are above a certain threshold in size.
In the usual case this approach takes place in a slow time scale,
in which events happen instantaneously (in comparison with
the waiting times, earthquakes are a good example).
When the threshold is high enough, the events can be considered
as extreme events, but there is no formal difference between these
and ordinary events.
%our approach is valid in any case.
Models of self-organized criticality are not very useful at this point, 
because, although the most common SOC models
lead to a Poissonian occurrence of events and therefore to
an exponential distribution of waiting times
(in contrast to some observational data \cite{Boffetta}),
other SOC models yield correlated avalanches \cite{Sanchez_prl,Davidsen_pre02} 
%CITAR!!. Joern PRE 2002? Sanchez? , mi comment?
One concludes from here that SOC is not about 
the temporal occurrence of events (in the slow time scale)
but about their size (or duration).

But it is interesting anyhow to stretch the notion 
of scale invariance to the waiting-time problem.
One deals now with two variables (size and waiting time), 
and in this case the scale transformation $\mathcal{T}$ is written as 
\begin{equation}
\mathcal{T}[f(x,y)] =  c f(x/a,y/b),
\end{equation}
with $f(x,y)$ a bivariate function, and $a,b,c > 0$, the scale factors
of the transformation.
When $a,b,c$ are greater than one the transformation dilates the function
in its three directions; 
for instance, taking $a=b=c=100$ takes a function in the scale of, let us say, meters
to the scale of centimeters.

The condition of scale invariance comes in this case from imposing
\begin{equation}
\mathcal{T}[f(x,y)] =  c f(x/a,y/b) = f(x,y),
\end{equation}
for all $a$.
So, we are looking for a fixed point for $\mathcal{T}$ and
the only solution \cite{Christensen_Moloney} can be written in the form
\begin{equation}
f^*(x,y) = \frac 1{x^\alpha} H\left(\frac x {y^\beta}\right)
\label{scaling_original}
\end{equation}
with $H$ an arbitrary univariate function called scaling function and 
$b$ and $c$ not free but given by
$ c=  1 / {a^\alpha}$ and $b = a^{1/\beta}$.
We will refer to this functional form of $f(x,y)$ as
a scaling law.
Note that for the univariate problem, described just by $x$, the scaling function 
has to be a constant, and then a power law is left.
In fact, a power law is a particular case of a scaling law, 
but not the opposite, 
and so it is useful to keep the distinction between 
both concepts if one is involved with multivariate functions.
The scaling law can be expressed in other,
alternative, equivalent forms, as for instance,
\begin{equation}
f^*(x,y) = \frac 1{x^\alpha} H_2\left(\frac y {x^{1/\beta}}\right)
= \frac 1{y^{\alpha\beta}} H_3\left(\frac y {x^{1/\beta}}\right),
\end{equation}
etc., just remember that the scaling function $H$ is arbitrary.

However, one can immediately realize that,
when events above a low threshold
in size are compared to events with higher size,
the analysis of the waiting-time 
does not only involve a scale transformation.
Indeed, an additional coarse-graining transformation is opperating.
This comes from the fact that events below the threshold need to be
eliminated, in order to measure the time between consecutive events
above the threshold. So, the transformation of the data comprises
a coarse-graining followed by a scale transformation \cite{Corral_prl.2005,Corral_jstat}.
This is the essence of a renormalization-group transformation
\cite{Christensen_Moloney}.

On the other hand, one can analyze SOC models and SOC-like systems
in a complementary way, not in the slow time scale in which almost instantaneous events happen but in the fast time scale inside of the events.
Then, one takes a threshold not in size but in intensity (size per unit time)
and measures the analogous of the waiting time, 
the time the signal is below threshold,
which in this context is called quiet time.
This approach was pursued in Ref. \cite{Paczuski_btw},
where a totally different behavior was found, 
with a power-law distribution of quiet times
(in agreement then with some observational results \cite{Boffetta}).
Naturally, this framework is valid not only for SOC-like systems
but for any time series from which events are defined 
by a process of thresholding.

This paper has as a goal 
to approach both kind of problems, 
waiting times defined in a slow time scale
and quiet times defined in a fast time scale,
from the point of view of scaling theory.
In the next section the renormalization-group approach
to the waiting-time sequence will be revisited
and the precise form of the resulting scaling law
will be presented for two broad class of processes, 
in analogy with the central limit theorem.
A different scaling form for the quiet-time distribution will be
used in the third section, generalizing the result for the diffusion of a Brownian
particle. This yields however important consequences
for the projection of extreme-event recurrence.
The result is extended to the interesting case in which the right tail
of the distribution is a power law, with such an exponent that the expected value of the variable is infinite.
The paper illustrates in which way one has to extrapolate the statistics of ordinary
events in order to infer the recurrence of extreme events
in a system that shows scaling.

\section{Scaling in marked point processes}

A point process, for our purposes, is a set of point events, i.e., instantaneous
events, that occur in time \cite{Cox_Isham,Daley_Vere_Jones,Lowen}. 
The simplest point process is the Poisson process, 
in which the events take place at a constant rate, 
independently of any other event.
A marked point process is a point process in which the points, 
i.e., the events, carry some ``mark'', which in our case 
corresponds to their size or dissipated energy.

As mentioned in the introduction, when one considers events
of a large enough size, the original point process 
transforms or renormalizes to a new point process.
Events below the threshold in size are eliminated whereas
events above the threshold survive. This coarse graining or decimation
is called thinning \cite{Cox_Isham} or rarefaction \cite{Gnedenko} in the theory of point processes.
As a special class of marked point processes we will consider 
in this section what is called marked renewal processes \cite{Daley_Vere_Jones}, 
in which the size of events
will be independent on other sizes and on the time of occurrence of the events, 
and additionally, the waiting times will be independent on previous waiting times
(and constitute then what is called a renewal processes \cite{Cox_Isham,Daley_Vere_Jones}, 
in which only the time of occurrence of the last event determines the time of the next one).

By construction, sizes are independent random variables, thus, the elimination of events
below the threshold in size is equivalent to a random thinning.
In this case, the waiting-time density transforms after thinning to
\begin{equation}
\tilde f_s(\omega) 
%=  p \tilde f(\omega) \sum_{j=1}^\infty [q \tilde f(\omega)]^{j-1} 
=\frac{p\tilde f(\omega)}{1- (1-p) \tilde f(\omega)}
\label{coarse}
\end{equation}
where $p$ is the probability of surviving to the thinning, 
i.e., the probability of being above the threshold $s$,
$\tilde f(\omega) $ is the Laplace transform of the original waiting-time 
probability density $f(t)$, and $\tilde f_s(\omega) $ is the Laplace transform
of the waiting-time density for events of size larger than $s$,
see Refs. \cite{Cox_Isham,Gnedenko,Corral_prl.2005,Corral_jstat}.
The Laplace transform arises because the waiting-time density for
events above the threshold is the convolution of a random number of
densities for all events. Then, going to Laplace space simplifies considerably
the equations, as convolutions transform into simple products there.

The subsequent step of the renormalization transformation
is a scale transformation. This can be done in two different
ways, a trivial one and a non-trivial one, 
as we will show in the next two subsections.

\subsection{Linear rescaling and trivial Poissonian fixed point}
\label{subsec_trivial}

 Let us considering the following scale transformation
for the waiting-time density for events above size $s$,
$f_s(t) \rightarrow p^{-1}f_s(t/p) $, or, in  Laplace space,
\begin{equation}
\tilde f_s(\omega) \rightarrow \tilde f_s(p \omega).
\label{linial}
\end{equation}
 This is because we expect that thinning increases the time scale in a $1/p$ factor,
and so, in order to compensate for this fact, 
we perform a contraction of the function in the $x-$axis with a scale factor $a=p < 1$,
whereas we perform a dilatation of the function in the $y-$axis 
with a scale factor $c=1/p > 1$, in order to keep normalization.
We will see in the next subsection that this makes sense
only if the mean waiting time $\langle t \rangle$ is finite.
The combined effect of the thinning (\ref{coarse}) and scale transformation yields the complete renormalization transformation $\mathcal{R}$, which is then
\begin{equation}
\mathcal{R} [\tilde f(\omega)] = \frac{p \tilde f(p \omega)}{1-q \tilde f(p \omega)}
\label{renorm} 
\end{equation}
with $q=1-p$. The solution of the fixed-point condition 
$\mathcal{R}[ \tilde f^*(\omega) ]= \tilde f^*(\omega)$, for all $p$,
leads to the fixed-point solution or renormalization-invariant Laplace
transform of the density $\tilde f^*(\omega)$.
Defining a new variable $u=p\omega$ and separating the variables $\omega$ and $u$
leads to the Laplace transform of an exponential distribution, 
i.e., 
\begin{equation}
\tilde f^*(\omega) = \frac 1 {1+\ell \omega}
\Rightarrow
f^*(t)=\frac 1 \ell e^{-t/\ell},
 \end{equation}
with $\ell$ an arbitrary constant, see Refs. \cite{Cox_Isham,Corral_prl.2005,Corral_jstat}.
This exponential distribution corresponds, due to the independence
of the events in the process, to the Poisson process.
In words, the Poisson process is 
invariant under the renormalization transformation given by Eq. (\ref{renorm}).
This is obvious if one intuitively knows the properties of the Poisson process.
But it is not only that the Poisson process is invariant under the transformation (\ref{renorm}),
the results also tells us that 
the Poisson process is the only marked renewal process invariant under such transformation
\cite{Cox_Isham,Corral_jstat}.
And, even more,
it is easy to show that it is also an attractor for any marked renewal process with waiting-time density with a finite mean (and whose Laplace transform exists), see Ref. \cite{Corral_jstat}.

This leads us to the first scaling form or data collapse considered in this paper.
Remember that in order to compare the thinned or coarse-grained distribution with the original one we perform a scale transformation 
of the form $f_s(t) \rightarrow p^{-1}f_s(t/p)$.
This means that the fixed-point solution, i.e., the
resulting probability density for very high thresholds in size
fulfills the following scaling law
\begin{equation}
f_s(t) = p F(p t), 
\label{scalinguno}
\end{equation}
where the scaling function $F$ is a decreasing exponential, 
but we will see that $F$ can take more general forms in other cases.
Therefore, if plotting $f_s(t)/p$ versus $pt$ yields a data collapse, 
i.e., a single curve, independently on the value of $p$,
this implies that the scaling law is fulfilled.

Other equivalent forms of the scaling law can be obtained
from the fact that the parameter $p$ gives the probability
that an event has energy $E$ larger than the threshold $s$
(given that has energy larger than a reference level 0), i.e., 
\begin{equation}
p = {\mbox{Prob} [E >s | E > 0]} \simeq \frac {N_s}{N_0}
\end{equation}
where $N_s$ is the number of events with energy above $s$.
Thus, the scaling $t \rightarrow p t$ is equivalent to
$
N_0 t \rightarrow N_s t
$
but this is only valid if the window width $T$ is fixed. If time windows of different width
are compared one needs to correct for this fact;
so, the rescaling is equivalent to
\begin{equation}
R_0 t = \frac{N_0}{T_0} t 
\rightarrow 
R_s t = \frac{N_s}{T_s} t
\end{equation}
where $R_s$ is the rate of events above $s$, 
i.e., the number of events above $s$ per unit time.
The situation of comparing different time windows is common
when dealing with real data of natural catastrophes, 
as different sizes have different
windows of completeness (see caption of Fig. 23 in Ref. \cite{Kanamori_rpp}).
Then, the scaling law (\ref{scalinguno}) can be alternatively written as
\begin{equation}
f_s(t) =R_s F( R_s t)
=\frac 1 {\langle t\rangle} F\left(\frac t {\langle t\rangle} \right),
\label{scalingdos}
\end{equation}
where we have introduced the mean waiting time
$\langle t \rangle = 1/R_s$ 
(which of course depends on the threshold $s$ although this is not reflected in the notation).
Note that the scaling function $F$ appearing here is not exactly the same   
as the one in Eq. (\ref{scalinguno}), as two multiplicative constants are reabsorbed in $F$.
In this case the scaling law can be fulfilled by plotting the dimensionless
waiting-time density $f_s(t)/R_s = \langle t \rangle f_s(t)$
versus the dimensionless time $R_s t = t/\langle t \rangle$
(in other words, plotting the waiting-time density in units of $R_s$
versus the waiting time in units of $\langle t \rangle$). 

In order to end with the different scaling forms, if there is scale invariance in the sizes (as it happens in SOC systems, in earthquakes, etc.), then $p\propto 1/s^\phi$, 
so, an additional form for the scaling law is
\begin{equation}
f_s(t) 
=\frac 1 {s^\phi} F\left(\frac t {s^\phi} \right),
\label{scalingtres}
\end{equation}
where again, some multiplicative constant has been
dropped deliberately.
This form of the scaling law is the ``most natural one''
in the sense that it is written in terms of the two variables of the problem, 
$t$ and $s$. The equivalence with the scale-invariant form of Eq. (\ref{scaling_original})
is achieved by realizing that $\alpha=1/\beta=\phi$.

These scaling laws, Eqs. (\ref{scalinguno}),  (\ref{scalingdos}), and  (\ref{scalingtres}),
are very useful in practice, as they have been shown to show up in several natural
hazards, as for instance, earthquakes \cite{Corral_prl.2004}. However, the scaling function
$F$ is not provided by an exponential function alone; 
rather, it contains a decreasing power-law for small to intermediate times,
with exponent $\nu<1$.
Other authors, using a simplified model of seismicity, have argued for a different behavior,
with no scaling law \cite{Saichev_Sornette_times}. 
In any case, the simple approach in this section, based in a process with independent times
and sizes, is clearly not valid for earthquakes and other natural hazards, 
in which important dependences exist \cite{Corral_tectono}.

Other examples of a general scaling law under thinning
with non-exponential scaling function include fractures
\cite{Davidsen_fracture,Baro_Corral}, forest fires \cite{Corral_fires}, and, curiously, printing requests \cite{Harder_Paczuski}.
In the case of fires only Eqs. (\ref{scalinguno}) and (\ref{scalingdos}) are valid, and  not Eq. (\ref{scalingtres}), as the fire size is not power-law distributed,
for the particular data analyzed in Ref. \cite{Corral_fires}.
The scaling is also found in a different approach for earthquakes
\cite{Bak.2002,Corral_pre.2003,Corral_physA.2004}, but there not only simple
thinning is performed and the scaling under thinning should arise from the underlying scaling explained in Ref. \cite{Corral_prl.2004}.
Other references provide only very indirect evidence of the scaling, 
as in nanofractures \cite{Astrom} and in tsunamis \cite{Geist_Parsons},
the problem being that no different thresholds in size are considered,
although the resulting scaling functions seem to be the same as for earthquakes.
The recurrence of words in texts, despite yielding a function very similar
to the earthquake case, has nothing to do with thinning or renormalization \cite{Corral_words}.
Finally, approaches in which the threshold is imposed not in size but in intensity
will be considered in the next section \cite{Bunde,Baiesi_flares,Laurson_upon},
whereas the approach of Ref. \cite{Yamasaki} is for a transformation of the original signal.

\subsection{Nonlinear rescaling and non-trivial Poissonian fixed point
\label{subsec_nontrivial}}

We explain in this subsection how the linear scaling with $p$ in 
Eq. (\ref{linial}) is not the only reasonable possibility.
Let us consider instead a constant $r>1$ 
and perform the rescaling as $f_s(t) \rightarrow p^{-r}f_s(t/p^r)$;
this yields the (completed) renormalization transformation $\mathcal{R}$,
using Eq. (\ref{coarse}),
\begin{equation}
\mathcal{R} [\tilde f(\omega)] = \frac{p \tilde f(p^r \omega)}{1-q \tilde f(p^r \omega)}.
\label{renormdos} 
\end{equation}
The fixed point condition leads in this case, in the same way as in the previous subsection, to 
\begin{equation}
\tilde f^*(\omega) = \frac 1 {1+\ell \omega^{1/r}}.
\label{Laplace_r}
\end{equation}
It is easy to show \cite{Corral_jstat} that, for $r>1$, this corresponds to a waiting time density
with a double power-law behavior, i.e.,
\begin{equation}
f^*(t) \propto
\left\{ \begin{array}{lll}
1/t^{1- 1/r}&\mbox{for}& t\ll \ell^{1/r}\\
1/t^{1+1/r}&\mbox{for}& t\gg \ell^{1/r},\\
\end{array}\right.
\end{equation}
which leads to an infinite mean, as the tail is a power law with an
exponent $1+1/r < 2$.
In the same way as for the trivial $r=1$ case,
this distribution is not only the only fixed point, for a fixed $r$,
but it is also an attractor for distributions with power-law tail, 
of the form $f(t)\sim B/t^{1+1/r}$. 
The case $r<1$ does not lead to well-defined probability distributions.

The non-linear rescaling  with $p^r$ is explained by the generalized
central limit theorem. Indeed, as the times $t_i$ have a distribution with 
a power-law tail, with exponent smaller than 2, 
the classic central limit theorem does not hold and one has instead
that the total time $T$ scales as
\begin{equation}
T=\sum_{i=1}^N t_i \propto N^r,
\end{equation}
see Ref. \cite{Bouchaud_Georges} or the Appendix I here.
Thus, after thinning by a factor $p$, the time window which restores the number
of events $N$ is the one that contains $N/p$ events, 
and this is the time window of duration $T/p^r$.
In other words, we need to rescale time as $t \rightarrow t/p^r$.
This is the origin of the non-linear rescaling.

In the same way as in the previous subsection,
invariance under the renormalization transformation 
means that the invariant density is of the form
\begin{equation}
f_s(t) = p^r G(p^r t),
\label{pr}
\end{equation}
with $G$ a scaling function given by the inverse Laplace transform of 
Eq. (\ref{Laplace_r}), which we have explained has
a double power-law form.

Expressions alternative to the previous rescaling can be
obtained as before, by using that
\begin{equation}
p^r \simeq \frac{N_s^r/T_s}{N_0^r/T_0} \propto \frac 1 {s^{\phi r}}
\end{equation}
where $N_s$ is the number of events with size (energy) above $s$
and we have taken into account that, in practice, 
the time window can be different for events above size $s$
than for events above the reference size $0$.

Therefore, the scaling law (\ref{pr}) can be written as
\begin{equation}
f_s(t) = 
\frac 1{\sum t_i/ N_s^r} G\left( \frac t {\sum t_i/ N_s^r}\right)
%%= \frac 1{\bar t_s N_s^{1-r} } G\left( \frac t {\bar t_s N_s^{1-r} }\right),
= \frac {N_s^{r-1}}{\bar t_s } G\left( \frac {N_s^{r-1} t} {\bar t_s }\right),
\label{specialuno}
\end{equation}
where $\sum t_i$ refers to events above threshold $s$
and $\bar t_s$ the sample mean of the waiting time above $s$
(different from the mean of the distribution, which is infinite).
Note that the generalized central limit theorem implies that
$\bar t_s$ diverges with the number of data, 
but $\sum t_i /N_s^r$ does not.
In the case $r=1$, then $\bar t_s \simeq \langle t \rangle$
and we recover the scaling of the previous subsection.
If we put the size-dependence explicitly,
\begin{equation}
f_s(t) =\frac 1 {s^{\phi r}} G\left(\frac t {s^{\phi r}}\right).
\label{specialdos}
\end{equation}
In this one %Eqs. (\ref{specialuno}) 
and in Eq. (\ref{specialuno}) we have deliberately
ignored some multiplicative constants.
Comparison with the scale-invariant solution of the scale transformation,
Eq. (\ref{scaling_original}), shows that $\alpha=1/\beta=\phi r$.

A marked renewal process of the sort of the one in this subsection
is neither implausible nor artificial.
Indeed, consider that a hidden signal triggers instantaneous events (as earthquakes)
when it reaches a fixed threshold and that
at that point the signal is immediately
reset just below the threshold.
If the sizes of the events are independent
and the signal cannot reach an absorbing boundary below it
(this can be achieved in practice by
putting the boundary of a random walk at $-\infty$, or by
 taking the exponentiation of the random walk,
which never reaches the absorbing boundary at zero), then
this leads to a marked renewal process with waiting-time density
given by the so-called L\'evy-Smirnov distribution, see Eq. (\ref{Levy_Smirnov}),
which has a right power-law tail with exponent 3/2 \cite{Redner}.
Iterative application of the thinning transformation leads to a different distribution, 
which keeps the power-law tail with exponent $3/2$
for large times but develops a new power law in the regime of small $t$,
with exponent 1/2.

\subsection{Paralellism with the generalized central limit theorem}
\label{subsec_renorm}

Several references have pointed out the relationship between the central
limit theorem and renormalization \cite{Jona_Lasinio,Sornette_critical_book,Sethna_book},
including the generalized case \cite{Calvo}.
From our perspective,
we can provide a clear connection between the renormalization
under random thinning explained in the previous subsections
and the generalized central limit theorem.
The key is to consider
% this relationship is totally clear.
a point process in which events are not removed randomly
but deterministically, surviving a proportion $p$, in such a way that
out of $1/p$ events, the $1/p-$th survives and the rest are removed
(with $1/p$ integer, for instance, if $p=1/2$ the events are removed alternatively).
Then, for a process in which the waiting times are independent
(i.e., a renewal process), this deterministic thinning transformation 
can be written as
\begin{equation}
f_s(t) = [f(t)]^{\star 1/p},
\end{equation}
where 
$f(t)$ is the probability density 
of the original point process, $f_s(t)$ the corresponding one
after thinning
(associated to events above size $s$),
and ${\star 1/p}$ denotes convolution $1/p$ times.
In Laplace space, the combined thinning plus rescaling transformation becomes
\begin{equation}
\mathcal{R}[\tilde f(\omega) ] = [\tilde f(p^r\omega)]^{1/p},
\end{equation}
which is the equivalent of Eq. (\ref{renormdos}),
and then, the fixed point condition for $\mathcal{R}$, taking logarithms,
fulfills
\begin{equation}
\ln \tilde  f^*(\omega) = \frac 1 p \ln \tilde f^*(p^r\omega)
\end{equation}
This means that $\ln \tilde  f^*(\omega)$, which is a cumulant generating function, 
has to fulfill the scale-invariance condition. If we impose the fixed point condition for all $p$ the solution is then a power law, 
\begin{equation}
\ln \tilde  f^*(\omega) = -A\omega^{1/r},
\end{equation}
and for the Laplace transform of the density we find,
\begin{equation}
\tilde f^*(\omega) = e^{-A \omega^{1/r}}.
\end{equation}
For the case $r=1$ we recognize the Laplace transform of a Dirac's delta function,
$f^*(t) =\delta(t-A)$,
whereas for $r=2$ we get
\begin{equation}
f^*(t) = e^{-A^2/(4t)} \frac A {2 \sqrt \pi \, t^{3/2}},
\label{Levy_Smirnov}
\end{equation}
which is sometimes called L\'evy-Smirnov distribution.
For any $r>1$ the fixed-point distribution has a power-law tail, for large
times, with exponent $1+1/r < 2$
(with infinite mean therefore).
As one can see in Appendix I, these distributions are also attractors, 
with different domains of attraction. The case $r=1$
yields the classic law of large numbers \cite{Feller},
in the form of the attractiveness of the Dirac's delta function,
whereas $r>1$ leads to a 
certain case of the generalized central limit theorem \cite{Bouchaud_Georges}.

\section{Scaling in the quiet times of time series}

The previous section was devoted to point processes in which the size
of the events was larger than some threshold.
From the point of view of SOC, we were looking at the system in the slow time scale.
Here we analyze the equivalent problem in the fast time scale (if there are two time scales), in order to see similarities and differences.
Our approach will be also valid for non-SOC systems, 
with only one time scale, as usual time series.
The procedure consists simply in introducing a threshold 
in the value of the signal (intensity or activity in the SOC language)
and define the quiet time as the time the signal is below threshold.
Several papers have used this approach before \cite{Bunde,Baiesi_flares,Laurson_upon}.
Nevertheless, the results will be of broader applicability.
%%QUE DARIA PARA WHITE NOISE??? % falta pendiente

\subsection{Non-parametric scaling form}

In Appendix II we consider an intensity signal modeled by a Brownian
noise between two absorbing boundaries (the zero-intensity state and the threshold)
and show how the quiet-time distribution fulfills a scaling law.
Let us generalize here that scaling law, replacing
the Brownian exponent 3/2 by a generic and undefined exponent $\nu$, 
and considering a generic scaling function $G$, then
we write
\begin{equation}
f_\tau(t) = \frac 1 {\tau^{\nu}} G\left(\frac t {\tau}\right),
\end{equation}
where the scale parameter is now called $\tau$,
and enters as a subindex of the probability density $f_\tau(t)$
to indicate the dependence on the threshold
(as $\tau$ depends on it).
The density is defined for $t>m$ and is zero otherwise.
The scaling function $G$ has to decrease, for large arguments, 
fast enough, which in our case means faster than $1/t^6$,
as we are interested, at most, in the calculation of the third moment
(i.e., we ask not only for the finiteness of the expected value of $t^3$, 
which is $\langle t^3\rangle$, but also for the finiteness of the second moment
of $t^3$, which is $\langle t^6 \rangle$).
This is the ``worst'' scenario and in some case $G$ can decay 
slower than $1/t^6$, the main requirement being that
the moments involved in the equations are finite, 
as well as their variances.

As shown in Appendix II, the scaling function $G$
can be a bit complicated and
it is more convenient to write the scaling law in a slightly modified
form, separating a power-law behavior for intermediate values
from the rest of the scaling function,
\begin{equation}
f_\tau(t) =
\frac K \tau \left(\frac \tau t\right)^\nu F\left(\frac t \tau \right)
\label{alphapeq}
\end{equation}
for $t > m$,
where we assume that the function $F$ tends to a positive constant 
at zero, and $K$ is a normalization constant.
This scaling form also comprises the scaling laws
of the previous section, taking $0 < \nu <1$
for the case of earthquakes \cite{Corral_prl.2004},
and $\nu=0$ for the trivial Poisson fixed point
of uncorrelated point processes.
As, for small arguments, the scaling function $F$ tends to a constant
and decreases ``fast enough'' for large arguments, 
the scale parameter $\tau$ separates at least two regimes,
being the one for $t\gg \tau$
the corresponding for extreme events;
in other words, $\tau$ sets the scale of extreme events.
It is worth mentioning that here we use a broader definition of
scaling function than the one of Ref. \cite{Christensen_EPJB},
as there it is enforced that the scale parameter $\tau$ only appears
in the argument of the scaling function, and not outside.
The use of that prescription or the use of our scaling form 
is a matter of choice and has no fundamental implications.
%%OJO 2 REGIMENES O 3??? % falta pendiente

Thus, in the case $\nu < 1$ we can take $m=0$ and verify that $K$ is a constant indeed
and then it is reabsorbed into the scaling function (i.e., $K=1$).
However, for $\nu > 1$ one finds that $K$ is not a true constant but
proportional to $(m/\tau)^{\nu-1}$,
and then $m$ cannot be zero.
The scaling law for $\nu>1$ turns out to be
\begin{equation}
f_\tau(t) =
\frac 1 m \left(\frac m t\right)^\nu F\left(\frac t \tau \right)
\mbox{ for } t > m > 0.
\label{alphagran}
\end{equation}
Both cases, $\nu < 1$ and $\nu > 1$, can be summarized into
a unique scaling law,
\begin{equation}
f_\tau(t) =
\frac 1 \theta \left(\frac \theta t\right)^\nu F\left(\frac t \tau \right)
\mbox{ for } t > m,
\label{scalingwithF}
\end{equation}
with $\theta=\tau$ and $m=0$ for $\nu < 1$ 
and
with $\theta=m$ and $m>0$ for $\nu > 1$. 
Note that for $\nu > 1$ we can write 
\begin{equation}
f_\tau(t) = \frac 1 m \left(\frac m \tau \right)^\nu G\left(\frac t\tau\right),
\label{scalingfinal}
\end{equation}
although for $\nu < 1$ we still have $f_\tau(t) = G(t/\tau)/\tau$.

The scaling law becomes more apparent substituting the scale parameter
$\tau$ by its scaling with the threshold $L$ (assuming it happens, 
as for a Brownian particle), i.e., $\tau \propto L^\phi$.
In this way
\begin{equation}
f_\tau(t) = \frac 1 {L^\phi} G\left(\frac t {L^\phi}\right),
\end{equation}
for $\nu < 1$, and
\begin{equation}
f_\tau(t) = \frac 1 m \left(\frac m {L^\phi} \right)^\nu G\left(\frac t{L^\phi}\right),
\end{equation}
for $\nu > 1$.

If one is interested in describing the distribution, instead of by its probability density, 
by its cumulative or survivor function, defined by
$S_\tau(t)=\int_t^\infty f_\tau(t')dt'$,
one has that
\begin{equation}
S_\tau(t) \propto 
\left\{\begin{array}{ll}
I(t/\tau)
 & \mbox{ for }  \nu < 1,\\
(m/\tau)^{\nu-1} I(t/\tau)
 & \mbox{ for } \nu > 1, \\
\end{array}\right.
\end{equation}
with $I$ a new scaling function.
Anyhow, for us the most useful quantity
will be the density, $f_\tau(t)$.

At this point it becomes interesting to see the consequences of the scaling
law on the moments of the distribution
(moments about the origin, $t=0$).
One can easily calculate the mean 
for $\tau \gg m$ \cite{Christensen_Moloney},
\begin{equation}
\langle t \rangle = \int_m^\infty t f_\tau(t) dt \propto 
%%\tau^{2-\nu} m ^{\nu-1} =
\left\{\begin{array}{ll}
\tau  & \mbox{ for }  \nu < 1,\\
m (\tau/m)^{2-\nu}
 & \mbox{ for } 1 < \nu < 2, \\
m  &\mbox{ for } 2 < \nu, \\
\end{array}\right.
\label{scalingt}
\end{equation}
and in the same way the second moment,
\begin{equation}
\langle t^2 \rangle = \int_m^\infty t^2 f_\tau(t) dt \propto 
%%\tau^{2-\nu} m ^{\nu-1} =
\left\{\begin{array}{ll}
\tau^2 & \mbox{ for }  \nu < 1,\\
m^2 (\tau/m)^{3-\nu}
 & \mbox{ for } 1 < \nu < 3, \\
m^2  &\mbox{ for } 3 < \nu, \\
\end{array}\right.
\label{scalingt2}
\end{equation}
The idea is that the scaling forms we have seen up to now are parametric, 
in the sense that depend on unknown parameters 
$\nu$ and $\tau$.
The exponent $\nu$ may be calculated with careful methods
\cite{Corral_Deluca}, but the scale parameter needs a particular parameterization
of the scaling function $G$, which may me arbitrary.
An alternative, free of these restrictions, is to use a non-parametric scaling form.

For the case $\nu < 1$ we know that $\tau \propto \langle t \rangle$
(see Eq. (\ref{scalingt}))
and then we recover the scaling form proposed in the previous section
\cite{Corral_prl.2004}, 
\begin{equation}
f_\tau(t)= \frac 1 {\langle t \rangle} G\left(\frac t {\langle t \rangle}\right),
\end{equation}
where the constant of proportionality between $\tau$ and $\langle t \rangle$
is reabsorbed in $G$.
We could also have used that $\tau \propto \sqrt{\langle t^2 \rangle}$,
but the scaling with $\langle t \rangle$ is preferred as the computation of the first moment
has a smaller error than that of the second moment.

For the new case $1<\nu < 2$ we cannot apply the previous scaling, and instead, we
have, on the one hand, dividing Eqs. (\ref{scalingt2}) and (\ref{scalingt}),
\begin{equation}
\tau \propto \frac {\langle t^2 \rangle }{\langle t \rangle },
\end{equation}
and, on the other hand, isolating from Eq. (\ref{scalingt})
and substituting in the previous one,
\begin{equation}
\left(\frac m \tau \right)^\nu
\propto \frac{\langle t \rangle}m \left(\frac m {\tau}\right)^2 \propto m \frac {\langle t \rangle^3}{\langle t^2 \rangle^2}.
\end{equation}
%$$
%\tau^\nu \propto m^{\nu-1} \frac {\tau^2}{\langle t \rangle} \propto m^{\nu-1}\frac{\langle t^2 \rangle^2}{\langle t \rangle^3}.
%$$
Substituting this into the scaling law (\ref{scalingfinal}) we get
\begin{equation}
f_\tau(t)=
\frac {\langle t \rangle^3} {\langle t^2 \rangle^2}\,
G\left(\frac {\langle t \rangle } {\langle t^2 \rangle } \, t\right),
\label{Rosso}
\end{equation}
see Refs. \cite{Rosso,Peters_Deluca}. Figure 1 shows an example of this rescaling.

\begin{figure}
\centering
\includegraphics*[height=0.40\textwidth]{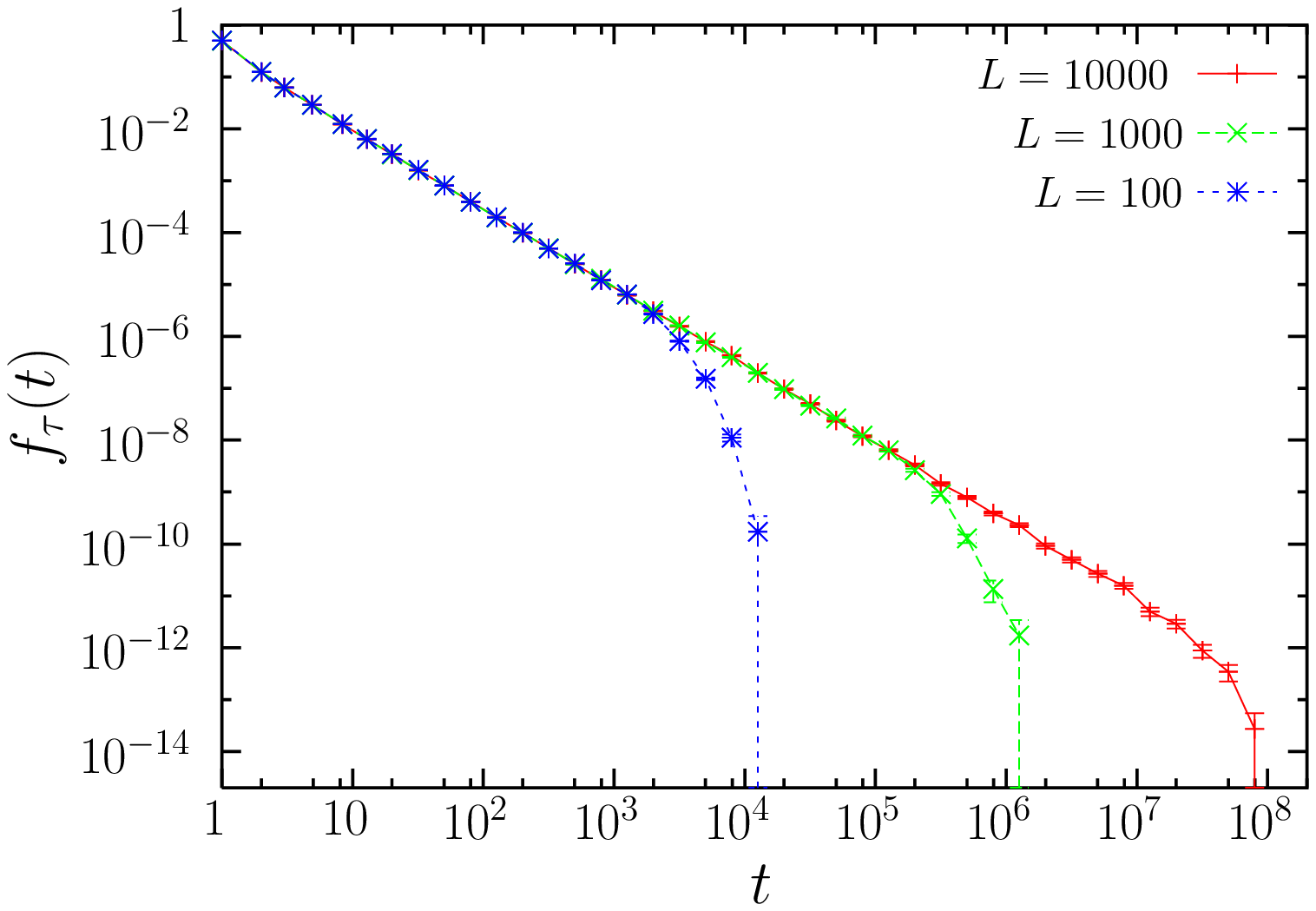} \\%}
\includegraphics*[height=0.40\textwidth]{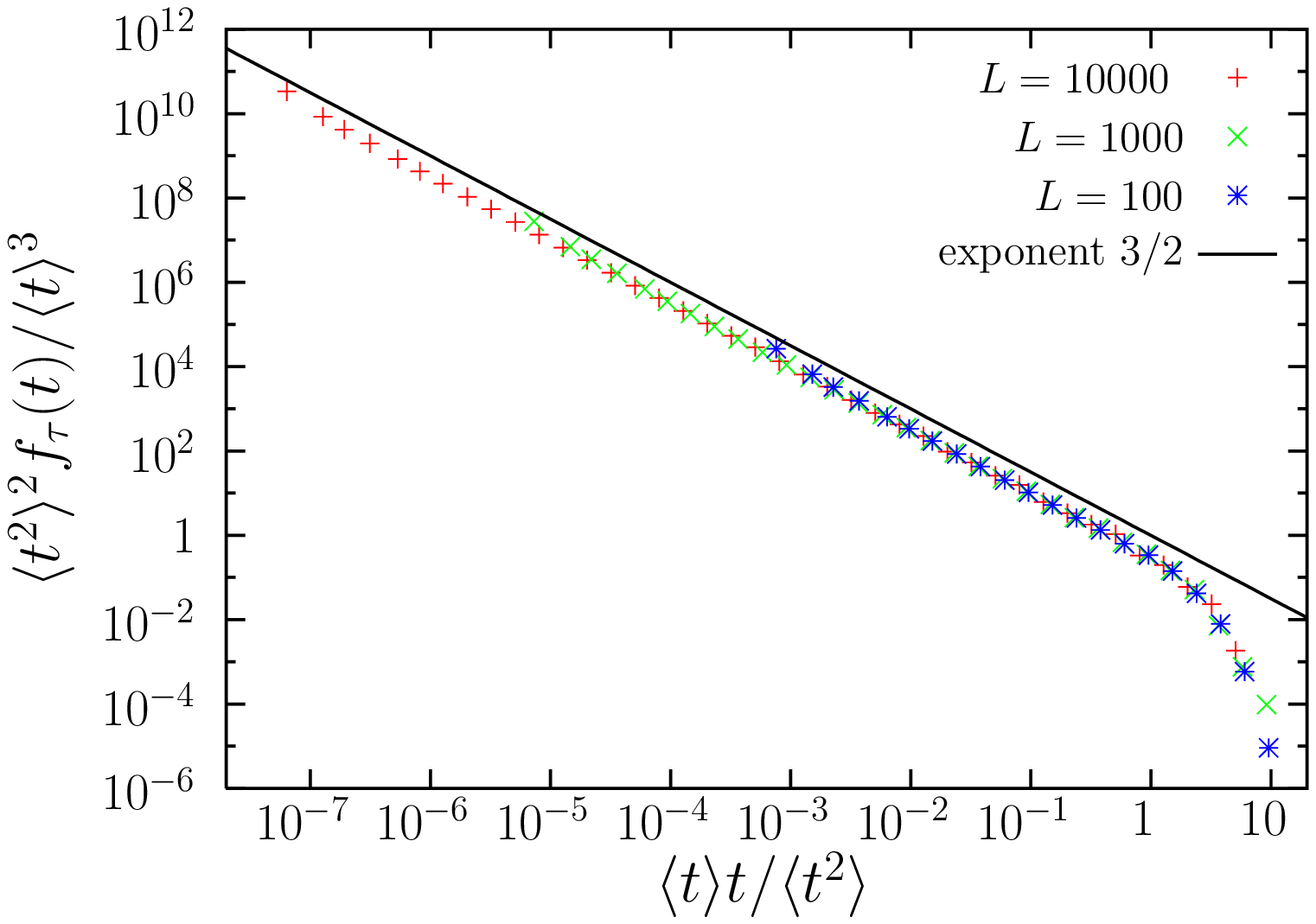}
 \caption{(a) Probability densities for the first-passage time of a random walk
between two absorbing boundaries.
Every time step the walker has probability 1/2 of increasing its position by 1
and the same probability to decrease it by 1.
The initial position is one unit below the threshold.
The time steps take place every 0.5 time units
(in order that the first-passage times take values 1, 2, 3, etc.)
Different values of the threshold $L$ are considered.
(b) Same distributions under the non-parametric rescaling given by
Eq. (\ref{Rosso}). The data collapse is an indication of the fulfillment of the scaling law.
}
\flabel{figuno}
 \end{figure}

An intermediate, semi-parametric scaling form is also possible, if one knows the exponent $\nu$
but not the scale parameter $\tau$,
in this case one may write, from Eq. (\ref{scalingt}),
%\begin{equation}
%f_\tau(t)=\frac 1 m \left(\frac{m \langle t \rangle}{\langle t^2 \rangle}\right)^\nu
%G\left(\frac {\langle t \rangle } {\langle t \rangle^2 } \, t\right) 
%\end{equation}
%$\tau \propto \langle t \rangle^{1/(2-\nu)}/m^{(\nu-1)/(2-\nu)}$
$\tau \propto m(\langle t \rangle/m)^{1/(2-\nu)}$
and from here,
\begin{equation}
%f_\tau(t)=\left(\frac {m^{2(\nu-1)}}{\langle t \rangle^\nu}\right)^{1/(2-\nu)}
%G\left(\frac {m^{(\nu-1)/(2-\nu)} } { \langle t \rangle^{1/(2-\nu)}} \, t\right) 
f_\tau(t)=\frac 1 m \left(\frac m {\langle t \rangle}\right)^{\nu/(2-\nu)}
G\left( \left (\frac m {\langle t \rangle}\right)^{1/(2-\nu)} \frac t m \right),
\end{equation}
see Refs.
\cite{Corral_hurricanes,Pruessner_comment}.

For completeness, it is interesting to analyze the scaling behavior for
$2 < \nu < 3$.
In this case the mean does not depend on the scale parameter, 
and we have to replace the mean with the third moment, 
which scales as $\langle t^3 \rangle \propto m^3 (\tau/m)^{4-\nu}$, 
valid for $1 < \nu < 4$.
Using also the scaling for $\langle t^2 \rangle$
and dividing both,
\begin{equation}
\frac{\langle t^3 \rangle}{\langle t^2 \rangle} \propto \tau,
\end{equation}
and isolating from the scaling of $\langle t^2 \rangle$,
\begin{equation}
\left(\frac m \tau\right)^\nu \propto \frac{\langle t^2 \rangle}{m^2}\left(\frac m \tau\right)^3,
\end{equation}
and substituting the scaling of $\tau$,
\begin{equation}
\left(\frac m \tau\right)^\nu \propto m \frac{\langle t^2 \rangle^4}{\langle t^3 \rangle^3},
\end{equation}
and substituting in the scaling law (\ref{scalingfinal}),
\begin{equation}
 f_\tau(t)=\frac{\langle t^2 \rangle^4}{\langle t^3 \rangle^3} G\left( \frac{\langle t^2 \rangle t}{\langle t^3 \rangle}\right).
\end{equation}
valid in principle for $1 < \nu < 3$ but
useful in practice for $2 < \nu < 3$.
A semi-parametric scaling form, 
if one knows $\nu$ but not $\tau$,
can be found from the fact that
$m/\tau \propto (m^2/\langle t^2\rangle)^{1/(3-\nu)}$
and then,
\begin{equation}
f_\tau(t) = \frac 1 m \left(\frac{m^2}{\langle t^2\rangle}\right)^{\nu/(3-\nu)}
G\left(  \left(\frac{m^2}{\langle t^2\rangle}\right)^{1/(3-\nu)}\frac t m\right).
\end{equation}
The different non-parametric scaling functions, for the different ranges of $\nu$,
are summarized in Table \ref{table_one}.

Scaling has important consequences for the projection of extreme events.
We have presented our scaling analysis for the waiting time or quiet time $t$
of events, but of course the results are more general, 
being suitable for other variables different than $t$.
Considering the scaling form (\ref{alphapeq}), valid for $\nu<1$,
we have seen that $\langle t \rangle \propto \tau$;
in consequence, an increase in the mean has associated a proportional increase
in the scale of extreme events, quantified by $\tau$.
But for the scaling form (\ref{alphagran}), linked to $\nu > 1$,
the mean does not scale linearly with the scale $\tau$ of extreme events.
In the case $1 < \nu < 2$ we have
$\langle t \rangle \propto \tau^{2-\nu}$ (see Eq. (\ref{scalingt}));
this means, for instance, that a two-fold increase in the mean of the distribution
implies an increase of a factor $2^{1/(2-\nu)}$ in the scale
of the largest, extreme events (a quadruplication if $\nu=3/2$
and much higher if $\nu$ approaches 2).
This is the case of the energy released by hurricanes \cite{Corral_hurricanes},
for which the mean of the distribution increases with sea-surface temperature,
but the increase of the largest energies is higher.
On the other hand, if $\nu > 2$,
the mean and the scale of extreme events become unrelated,
as $\langle t \rangle \propto m$ (i.e., a constant).
As a recent publication states, {\it ``much of today's research
represents climate change in terms of
changes to the annual or seasonal averages''},
whereas the subject of major concern is the change
corresponding to the most devastating, extreme events
\cite{Perez_ngeo}.

%\begin{widetext}

\begin{table}[ht]
\caption{Non-parametric rescaling of the axes coming from the scaling law
$f_\tau (t) =\theta^{-1} ( \theta /t)^\nu F(t /\tau)$
with $\theta=\tau$ for $\nu < 1$ 
and
with $\theta=m$  for $\nu > 1$.
The scaling function $F$ decays fast enough in order that
the moments up to $\langle t^3\rangle$ and their variances
are not infinite.
The equivalent rescaling for the survivor function 
$S_\tau(t)=( \theta /\tau)^{\nu-1} I(t /\tau)$
is also included.
}
\begin{tabular}{ l | c | c | c |}
& abscissa & ordinate for $f_\tau (t) $ & ordinate for $S_\tau (t) $ \\
\hline
$\nu < 1 $ & $t/\langle t  \rangle$ &  $\langle t  \rangle f_\tau(t)$ & $S_\tau(t)$\\
$1< \nu < 2 $ & $ \langle t  \rangle t/\langle t^2  \rangle$ 
&  $\langle t ^2 \rangle^2 f_\tau(t)/\langle t  \rangle^3$ 
&  $\langle t ^2 \rangle S_\tau(t)/\langle t  \rangle^2$ \\
$1< \nu < 3 $ & $ \langle t^2  \rangle t/\langle t^3  \rangle$ 
&  $\langle t ^3 \rangle^3 f_\tau(t)/\langle t^2  \rangle^4$ 
&  $\langle t ^3 \rangle^2 S_\tau(t)/\langle t^2  \rangle^3$ \\
\hline
 \end{tabular}
 \label{table_one} 
\end{table} 
%\end{widetext}

\subsection{Case with infinite mean generalized}

In the previous section (subsec. \ref{subsec_nontrivial}) 
we saw how, in some peculiar cases,
marked renewal processes do not converge under thinning to the 
Poisson process but to a process in which the waiting-time density
shows a double (decreasing) power-law behavior, with an exponent
between 0 and 1 for short times, and another exponent between 1 and 2
for long times (and with the sum of both exponents equal to 2).
This behavior in the right tail is associated to an infinite mean-waiting time 
$\langle t \rangle$.

In this subsection we study the proper scaling of such distributions
in a more general framework, considering distributions
(no matter if for the waiting time, the quiet time, or any other variable
showing scaling)
as those given by  Eqs. (\ref{scalingwithF}) and (\ref{scalingfinal}),
\begin{equation}
f_\tau(t) =
\frac 1 \theta \left(\frac \theta t\right)^\nu F\left(\frac t \tau \right)
= \frac 1 \theta \left(\frac \theta \tau \right)^\nu G\left(\frac t\tau\right),
\label{eqFG}
\end{equation}
defined for $t>m,$
with $m=0$ and $\theta=\tau$ if $\nu < 1$ 
and
with $m>0$ and $\theta=m$ if $\nu > 1$. 
Note that $t$ is the variable and $\tau$ is the scale parameter.
The difference with the previous subsection is that
$f_\tau(t)$ has now a (right) power-law tail, with exponent between 1 and 2,
i.e., for large arguments $G$ decays as a power law 
with exponent $1+\rho$, with $0 < \rho < 1$, and then $\langle t\rangle$ is infinite,
as in subsec. \ref{subsec_nontrivial}.
But in contrast to that subsection, and 
in the same way as in the previous one, $F$ goes to a constant for small arguments, 
so $G$ goes as another (decreasing) power law there, with exponent $\nu$,
with no particular relation between both exponents ($1+\rho$ and $\nu$).

In the same way as before, we are interested in relating the mean value of the variable with
its scale parameter $\tau$, with the difference that now 
the expected value $\langle t \rangle$ is infinite
and we have to emphasize the role of the sample mean, $\bar t =\sum_i t_i/N$
(which is always finite if $N$ is finite).
The power-law tail of the distribution means that $f_\tau(t)$
goes, for large enough $t$ as $B/t^{1+\rho}$.
We can realize that a scale parameter is given by $\tau=B^{1/\rho}$.
Applying the results of the generalized central-limit theorem explained in Appendix I, 
we get that
\begin{equation}
\sum_{i=1}^N t_i \sim [-\Gamma(\rho) B N]^{1/\rho},
\end{equation}
where the symbol ``$\sim$'' means that, for large $N$, $\sum_i t_i$ is distributed following a
certain (L\'evy-stable) distribution whose scale parameter is $[-\Gamma(\rho) B N]^{1/\rho}$.
From here we get that the scale parameter $\tau$ of $f_\tau(t)$
can be related to $\sum_i t_i$ and therefore with the sample mean as
\begin{equation}
\tau = B^{1/\rho}\sim
\frac{\sum_{i=1}^N t_i }{[-\Gamma(\rho)  N]^{1/\rho}}
\propto \frac {\bar t} {N^{1/\rho-1}}.
\end{equation}

The case $\nu < 1$, for which $\theta=\tau$, 
can be obtained straightforwardly from Eq. (\ref{eqFG}), 
which leads to 
\begin{equation}
f_\tau(t) = \frac{N^{1/\rho-1} }{\bar t} G\left(\frac{N^{1/\rho-1} t}{\bar t}\right)
\end{equation}
and is of the same form as Eq. (\ref{specialuno}).
%
%OJO, ALLI ERA CON F NO CON G  !!!! Arreglao, no??
%
For the new case $\nu >1$ we need to consider $\theta=m$, 
and from Eq. (\ref{eqFG}), substituting $\tau$, we get
\begin{equation}
f_\tau(t) = \frac 1 m \left(\frac{m N^{1/\rho-1} }{\bar t}\right)^\nu
G\left(\frac{N^{1/\rho-1} t}{\bar t}\right).
\label{supersuperscaling}
\end{equation}
Table \ref{table_two} summarizes the results for $\nu < 1$ and $\nu > 1$.

%\begin{widetext}

\begin{table}[ht]
\caption{Semi-parametric rescaling of the axes coming from the scaling law
$f_\tau (t) =\theta^{-1} ( \theta /t)^\nu F(t /\tau)$
with $\theta=\tau$ for $\nu < 1$ 
and
with $\theta=m$  for $\nu > 1$. The right tail of $F$ decays slow enough
in order that $\langle t \rangle$ is infinite.
In contrast, the sample mean $\bar t$ is finite if $N$ is finite.
}
\begin{tabular}{ l  | c | c |}
& abscissa & ordinate \\
\hline
$\nu < 1 $ & $N^{1/\rho-1} t/ \bar t $ &  $\bar t f_\tau(t)/N^{1/\rho-1}$ \\
$\nu > 1 $ & $N^{1/\rho-1} t/ \bar t $ &  $(\bar t)^\nu f_\tau(t)/N^{\nu(1/\rho-1)}$ \\
\hline
 \end{tabular}
 \label{table_two} 
\end{table} 
%\end{widetext}

Notice that, in contrast to the case of previous subsection, 
where the tail of the distribution was decaying fast enough,
we do not arrive at non-parametric forms of the scaling laws.
We can get ride of the scale parameter $\tau$ but
the results still depend on the parameter $\rho$ and, 
if $\nu >1$, also on $\nu$.
The reason is that, for fixed number of data, the (sample) moments do not depend 
on the exponents, and therefore the exponents cannot be isolated from the moments.
Indeed, in order to calculate the second sample moment $\bar{t^2}$
we can use that $t^2$ is also power-law distributed with exponent $1+\rho/2$
(and a multiplying constant $B/2$)
and we can apply in the same way the generalized central-limit theorem to get
\begin{equation}
\sum_{i=1}^N t_i^2 \sim \left[-\Gamma(\rho/2) \frac B 2 N\right]^{2/\rho},
\end{equation}
and from here
\begin{equation}
\tau^2=B^{2/\rho} \sim \frac{\bar{t^2}}{N^{2/\rho-1}}.
\end{equation}
For higher order moments the results are analogous, 
and so we cannot isolate neither the exponent $\nu$ nor $\rho$
from here (for a fixed $N$).

Both forms of the scaling law, for $\nu < 1$ and for $\nu > 1$,
can be tested from simulations.
We generate double-power-law distributed values of $t$
by the rejection method, first generating a (simple) power law
with exponent $1+\rho$ for $t>1$ and then rejecting the value of
$t$ if it yields a $t^\epsilon/(k + t^\epsilon)$ smaller than a uniform
random number $u$. This leads to a double-power-law distribution, 
with scale parameter $\tau=k^{1/\epsilon}$ and
with exponents $\nu=1+\rho-\epsilon$ and $1+\rho$
for small and large $t$, respectively (if $\epsilon >0$, if not, the situation is reversed).
Figure \ref{figdos} shows an example for $\nu>1$.

\begin{figure}
\centering
\includegraphics*[height=0.40\textwidth]{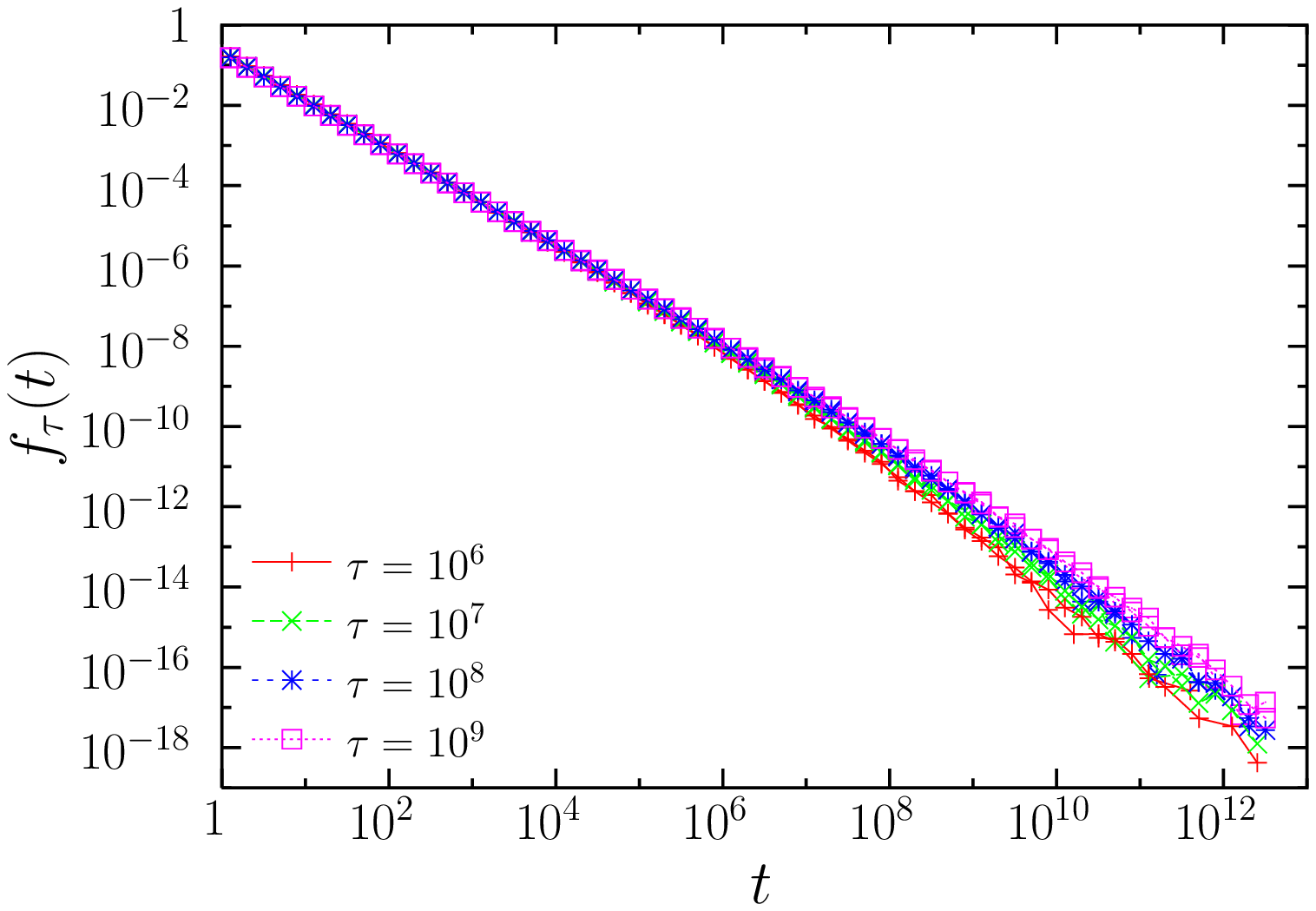} \\%}
\includegraphics*[height=0.40\textwidth]{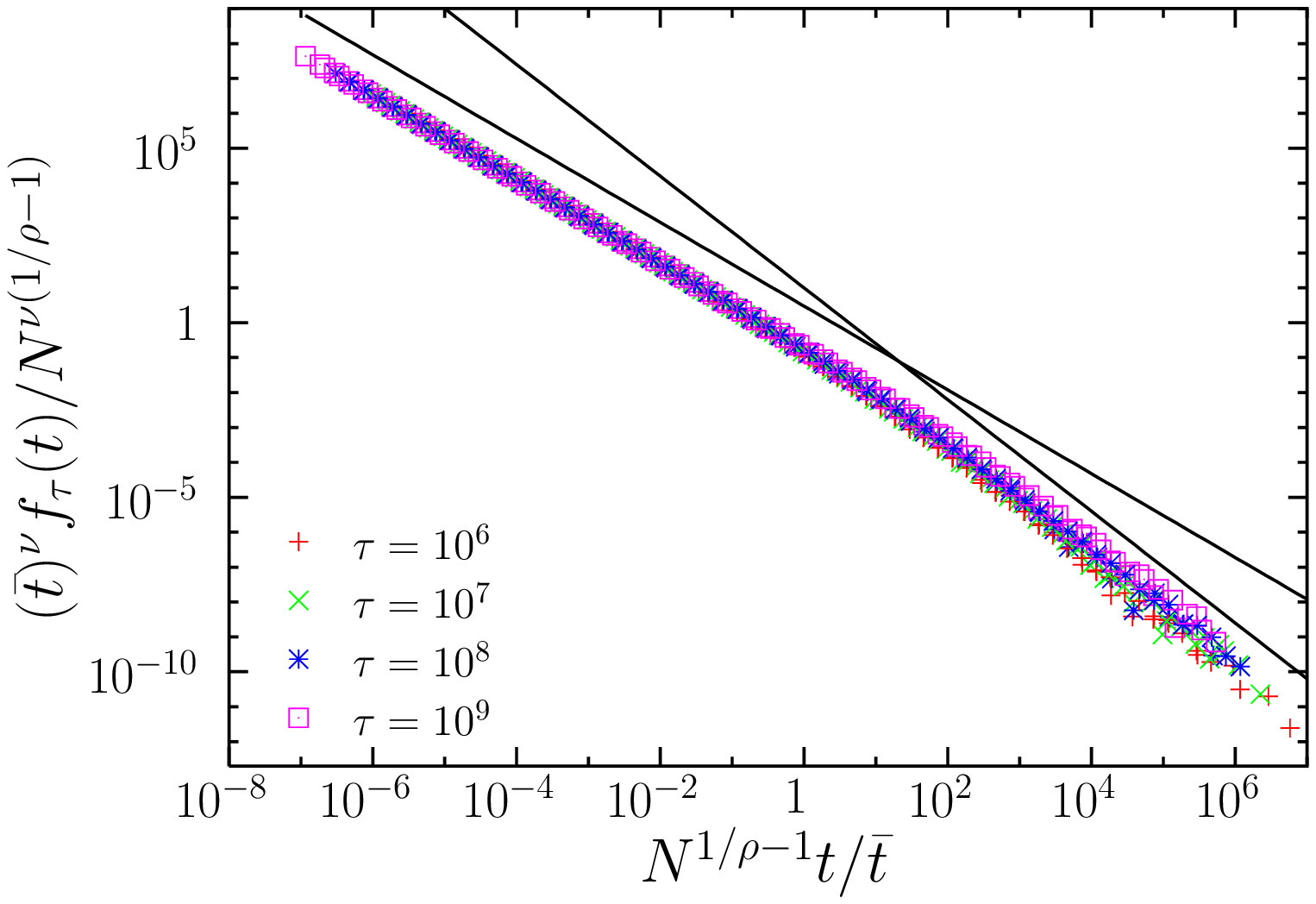}
 \caption{(a) Probability densities for simulated double-power-law distributions, 
with parameters $\nu=1.2$, $1+\rho=1.6$, and $m=1$, for different values of the scale
parameter $\tau$. Two different values of $N$ are used for each $\tau$, these are $N=10^5$
and $N=10^6$. (b) Same distributions after a semi-parametric rescaling of the axes, following Eq. (\ref{supersuperscaling}). The collapse of data indicates the fulfillment of the scaling law.
The straight lines are power laws with exponents 1.2 and 1.6.}
\flabel{figdos}
 \end{figure}

%%CITAR roc IN PREPARATION!!! % falta pendiente

\section{Summary}

We have seen how scaling laws can arise in the waiting time or quiet time
between extreme events. % (and not so extreme) events.
We consider two scenarios.
The first one deals with point processes, 
for which events happen instantaneously
in comparison with the waiting time between events.
This corresponds to the slow time scale of driving in SOC systems.
It is shown that scaling might arise as an attractive fixed point of a 
renormalization transformation, consisting of thinning of events below a threshold
in size plus a rescaling of the time axis (so, it is considered that high enough thresholds 
lead to the fixed-point solution).
For processes without correlations, the resulting scaling law 
contains an exponential scaling function (characteristic of a Poisson process)
if the mean waiting time between the original events has a finite mean.
In contrast, if the mean waiting time is infinite, a non-trivial scaling function
arises, with a double-power-law behavior.
Different forms of the scaling law are shown in terms of the probability
$p$ of surviving the thinning process, the mean waiting time for the renormalized
events, or the distribution of events above size $s$.
The results are critically dependent on whether $\langle t \rangle$ is finite or infinite.
As a corollary, a slight change in the renormalization procedure
(removing events deterministically instead of randomly)
leads to a direct derivation of the law of large numbers and the generalized
central-limit theorem (for the case of maximum asymmetry) \cite{Gnedenko}.

The second scenario contemplates time series, 
and the variable of interest is the quiet time
(the time the signal is below a threshold), 
representing again but in a different way the time 
between extreme events.
In the SOC framework this would correspond
to look at the system in the fast time scale inside the avalanches.
In this case scaling laws might show up 
from the first-passage time of diffusion processes
that model the signal, in the limit of high enough thresholds.
The scaling law yields scaling functions different than the ones
from point processes, with a decaying power law for short times
(except for the shortest times) and a much faster decay for long times.
Using the dependence of the moments of the distribution on the scale parameter
and on the power-law exponent we can replace the latter ones in terms of the former
ones and obtain a non-parametric scaling law, depending only on the moments.
The particular equation for the scaling law depends on the value of the power-law
exponent; in particular we have distinguished $\nu < 1$ (trivial scaling),
$1 < \nu < 2$, and $1 < \nu < 3$.
Finally, this result is generalized for the case in which the waiting-time distribution
does not decay fast enough for long times but slow enough in such a way
that the mean waiting time becomes infinite.
No non-parametric scaling is possible then and one obtains a semi-parametric
scaling in which the dependence on the scale parameter can be substituted by the 
dependence on the sample moments, but the exponents cannot be replaced.
Additionally,
important lessons for the projection of extreme events are derived from the different
scaling of the mean of the distribution and the scale of extreme events.

%\Acknowledgements

\section{Acknowledgements}

This paper is a re-elaboration of a talk of the author at the EXEV14 workshop;
invitation by A. Deluca, H. Kantz and N. W. Watkins is gratefully acknowledged.
The author is indebted to N. R. Moloney for pointing to the references
relating renormalization and the central limit theorem
and to G. Pruessner for his influential lectures at the
Joint CRM-Imperial College School in Complex Systems.
Research expenses
were founded by projects FIS2009-09508, from the disappeared Spanish
MICINN, FIS2012-31324 from Spanish MINECO, and 2009SGR-164, from AGAUR. 
This work is also associated to 2014SGR-1307.

\section{Appendix I: Generalized central limit theorem from renormalization}

Remember that in the (generalized-) central-limit-theorem scenario
one sums independent identically distributed random variables
and then the sum is rescaled in order that the resulting distribution
is comparable to the original distribution of the individual variables.
We showed in subsec. \ref{subsec_renorm}
that this is exactly the same problem one finds for the waiting time of extreme 
events when these happen in a deterministic way, 
being the $1/p-$th out of $1/p$ events an extreme event, 
and the rest, $1/p-1$ events, being ordinary events, which are removed.
In this way, the connection with a renormalization transformation is direct.
Introducing $n=1/p$, and assuming $n$ is a natural number,
we know from subsec. \ref{subsec_renorm} that the cumulant generating function of the sum and 
rescaling of these $n$ random variables (in our case waiting times, 
but we put ourselves here in a more general context) transforms as
\begin{equation}
\mathcal{R}[\ln \tilde  f(\omega)] = n \ln \tilde f\left(\frac \omega {n^r}\right),
\end{equation}
i.e., suffers a simple scale transformation under this operation,
where $\ln \tilde  f(\omega)$ is the cumulant generating function of the
original random variables (which is the logarithm of the Laplace transform $ \tilde  f(\omega)$
of the probability density $ f(t)$).
In  subsec. \ref{subsec_renorm} we saw the fixed points of this transformation
(i.e., the distributions which are stable under renormalization)
and now we explore the attractiveness of these fixed points,
which is what provides the content for the central limit theorem.

\subsection{Existence of all moments}

The simplest case corresponds to the common situation of an initial probability density 
whose moments exist and are finite (non-infinite).
If the cumulant generating function exists, it is well know that
its expansion will be
\begin{equation}
\ln \tilde  f(\omega) =-\langle t \rangle \omega + \mathcal{O}(\omega),
\end{equation}
where $\mathcal{O}(\omega)$ refers to terms that are of higher order than linear in $\omega$
(just expand by Taylor the exponential factor $e^{-\omega t} $ in the Laplace transform,
take the logarithm of the resulting series and expand again the logarithm).
%%, which works if all moments exist).
Applying the renormalization transformation to the expansion of $\ln \tilde  f(\omega) $
one gets
\begin{equation}
\mathcal{R}[\ln \tilde  f(\omega)] = n \left(-\langle t \rangle \frac\omega {n^r} + 
\mathcal{O}(\omega/n^r)\right),
\end{equation}
and imposing linear rescaling, $r=1$, leads, after successive iterations of the transformation
(or, equivalently, after taking the limit $n\rightarrow \infty$) to
\begin{equation}
\mathcal{R}[\cdots \mathcal{R}[\ln \tilde  f(\omega)]\cdots] \rightarrow -\langle t \rangle \omega,
\end{equation}
where note that all non-linear terms vanish
(we are assuming that $\langle t \rangle > 0$).
We already saw that this is the logarithm of the Laplace transform of a Dirac's delta function, 
and therefore
\begin{equation}
\mathcal{R}[\cdots \mathcal{R}[ f(t)]\cdots] \rightarrow \delta(t-\langle t \rangle)
\end{equation}
(we use the same notation for the renormalization transformation operating 
over $f(t)$, over its Laplace transform, or over the cumulant generating function).
Although we are in the context of the central limit theorem, this trivially-looking result
is nothing else than a weak version of the law of large numbers \cite{Feller}, 
perhaps the most fundamental result in probability theory and statistics,
ensuring that the arithmetic mean of a sample converges
to the expected value of the distribution (if this expected value exists and is not infinite, 
and the same for the rest of the moments). 

It is noteworthy that the equivalent of the law of large numbers when 
the number of variables that are added is not fixed but random, following a
geometrical distribution, is the exponential distribution
(as we saw in subsec. \ref{subsec_trivial}).
On the other hand, if a fixed number of variables are summed
but the mean is zero one obtains the classic central-limit result
in the form of a Gaussian distribution, which is of no interest here, 
but see Ref. \cite{Bouchaud_Georges}.

\subsection{Power laws and power-law tails}

The ``non-trivial''
case corresponds to a distribution $f(t)$ which is a power law, 
or at least that has a power-law upper tail.
In the first case $f(t) = B/t^{1+\rho}$, 
for $t>m>0$, with $\rho>0$ and $B=\rho m^\rho$
(normalization condition).
The Laplace transform of the density can be expressed
in terms of the incomplete gamma function,
\begin{equation}
\tilde f(\omega) = B \int_m^\infty e^{-\omega t} t^{-\rho -1} dt =
B \omega^\rho \Gamma(-\rho,m\omega).
\label{fomega}
\end{equation}
Using the expansion of this function (see 6.5.29 of Ref. \cite{Abramowitz}),
\begin{equation}
\Gamma(\gamma,z) = \Gamma(\gamma) - z^\gamma \sum_{n=0}^\infty 
\frac {(-z)^n}{(\gamma+n)n!},
\end{equation}
valid for $\gamma \ne 0, -1, -2, -3 \dots$, we can write
\begin{equation}
z^{\rho}\Gamma(-\rho,z) = z^\rho\Gamma(-\rho) - \sum_{n=0}^\infty 
\frac {(-z)^n}{(n-\rho)n!} 
\end{equation}
\begin{equation}
=\frac 1 \rho \left[ \rho \Gamma(-\rho)z^\rho + 
\left(1 + \frac{\rho z}{1-\rho} + \mathcal{O}(z) \right)\right],
\end{equation}
for $\rho \ne 0, 1, 2, \dots$
From here, taking the logarithm, 
% for small $z$,
and using its expansion around $z=1$,
\begin{equation}
\ln \left[ z^{\rho}\Gamma(-\rho,z) \right]
%=  -\ln \rho + \ln \left[ \, \right] 
= -\ln \rho +\rho \Gamma(-\rho)z^\rho  + \frac{\rho z}{1-\rho} +\mathcal{O}(z),
\end{equation}
which we can substitute in the expression of $\ln \tilde f(\omega)$, 
from Eq. (\ref{fomega}), so
\begin{equation}
\ln \tilde f(\omega) = \ln \frac B {m^\rho} + \ln\left[ m^\rho \omega^\rho \Gamma(-\rho,m\omega) \right]
\end{equation}
\begin{equation}
= %%ln B - \rho\ln m -\ln \rho +
\rho \Gamma(-\rho)m^\rho \omega^\rho  + \frac{\rho m \omega}{1-\rho} +\mathcal{O}(\omega),
\label{rhogrho}
\end{equation}
where we have used that $B=\rho m^\rho$.
Applying the renormalization transformation
we arrive at
\begin{equation}
\mathcal{R}[\ln \tilde  f(\omega)] = n \ln \tilde f\left(\frac \omega {n^r}\right)
\end{equation}
\begin{equation}
=n \left( \rho \Gamma(-\rho)m^\rho \left(\frac \omega {n^r}\right)^\rho  + \frac{\rho m \omega/n^r}{1-\rho} +\mathcal{O}(\omega/n^r)
\right).
\end{equation}

We can distinguish two situations here.
On the one hand, 
if $\rho > 1$ in order to have a well-defined limiting distribution we need to take
$r=1$, which yields, in the same way as in the trivial case, 
the delta function corresponding to the law of large numbers, 
\begin{equation}
\mathcal{R}[\dots \mathcal{R}[
\ln \tilde  f(\omega)]\dots] \rightarrow 
-\frac{\rho m \omega}{\rho-1},
\end{equation}
so, 
\begin{equation}
\mathcal{R}[\dots \mathcal{R}[  f(t)]\dots] \rightarrow 
\delta\left(t-\frac{\rho m}{\rho-1}\right).
\end{equation}
Notice that for the power-law distribution with $\rho > 1$
the expected value is
$\langle t \rangle = \frac{\rho m}{\rho-1}$,
but the other moments can diverge and still
the law of large numbers holds in this case.

On the other hand, a different situation corresponds to $\rho < 1$,
and in order to have a well-defined limit we need to take
$r=1/\rho$, and then
\begin{equation}
\mathcal{R}[\dots \mathcal{R}[
\ln \tilde  f(\omega)]\dots] \rightarrow 
\rho \Gamma(-\rho)m^\rho \omega^\rho,
\end{equation}
which corresponds to a Laplace transform given by 
$\tilde f^*(\omega)=\exp(B  \Gamma(-\rho) \omega^\rho)$.
This yields % a special case of the well known 
L\'evy-stable distributions
with maximal asymmetry \cite{Bouchaud_Georges}.
As far as the author knows, the Laplace transform only can be inverted
for $\rho=1/2$, yielding the sometimes called L\'evy-Smirnov distribution, 
Eq. (\ref{Levy_Smirnov}),
%\begin{equation}
%f^*(t) = e^{-A^2/(4t)} \frac A {2 \sqrt \pi \, t^{3/2}}
%\end{equation}
with $A=-B\Gamma(-1/2)=2\sqrt{\pi} B$.
For general values of $\rho$ (in the range $(0,1)$)
it can be shown that the limiting distribution behaves, 
for large $t$ as $f^*(t) \propto 1/t^{1+\rho}.$
So, when power-law distributed variables, with exponent $1+1/r<2$, 
are summed, the way to rescale them is not by the number of terms $n$
but by $n^r$. With this ``strange'' average we get convergence 
to a L\'evy stable distribution, 
but not to a Dirac's delta distribution, 
so, no generalized form of the law of large numbers holds.
Rather, this result is considered to belong to the generalized
central-limit theorem, although one is not dealing 
with the center of the distribution.

In the second ``non-trivial'' case we have a distribution which asymptotically (for large $t$) 
is a power law, 
$f(t) \sim B/t^{1+\rho}$, with $B$ not necessarily equal to $\rho m^\rho$.
Still it is possible to use the expansion of the Laplace transform of $f(t)$, which is
\begin{equation}
\tilde f(\omega) =B \Gamma(-\rho) \omega^\rho +\sum_{n=0}^\infty \frac {a_n (-\omega)^n}{n!},
\end{equation}
see 4.6.23 of Ref. \cite{Bleistein}. 
%% CONSULTAR!!!! % falta pendiente
As $a_0=1$ then, taking the logarithm,
\begin{equation}
\ln \tilde f(\omega) =B \Gamma(-\rho) \omega^\rho -  {a_1 \omega}+\mathcal{O}(\omega),
\end{equation}
and therefore we are in the same situation as for the pure power-law case, 
Eq. (\ref{rhogrho}).
The convergence is towards $f^*(t)=\delta(t-a_1)$ if $\rho>1$
(so, the law of large numbers holds)
and towards $\tilde f^*(\omega) = \exp(-A \omega^\rho)$
if $\rho<1$, with $A=-B\Gamma(-\rho)$
(notice that $A>0$ and that the scale parameter of $f^*(t)$ is $A^{1 /\rho}$).
This justifies the non-linear rescaling of the sum of $n$ random variables, 
so the sum ``goes'' as  $(A n)^{1/\rho}$
but it is broadly distributed.
In other words, $\sum t_i / n^r$ follows a L\'evy-stable distribution
(with maximal asymmetry) with scale parameter
$[-\Gamma(\rho) B]^{1/\rho}$.

\section{Appendix II: Quiet time of a Brownian signal}

In order to understand how scaling can show up
in the times a time series is below a threshold, 
let us consider that the signal is given by the position
of a Brownian particle diffusing between two absorbing boundaries. 
The particle starts below but very close to the threshold, which is an absorving boundary
(assuming it had just crossed the threshold from above),
and yields a quiet time if it reaches the threshold;
on the contrary, if it reaches the zero-value intensity, the process
(the SOC avalanche) dies out and no quiet time is computed.
The quiet times generated in this way define a renewal process, 
in which the distribution of quiet times completely describes the process.
A description of quiet times in the most complete framework 
would involve the use of copulas \cite{Chicheportiche_copulas}.

%%ESCRIBIR a REDNER!! % falta pendiente

The first step is the solution of the diffusion equation, 
$\frac{\partial C}{\partial t}=D \frac{\partial^2 C}{\partial x^2}$,
with $D$ the diffusion coefficient.
For simplicity we take the $x-$axis in such a way that the intensity threshold
corresponds to $x=0$, whereas the other absorbing boundary, 
at zero intensity, corresponds to $x=L$
(in this way, increasing intensity corresponds to decreasing $x$).
In any case, $L$ is the separation between the threshold 
and the zero-intensity absorbing boundary. 
It is well known \cite{Crank_diffusion,Redner} that the solution to this diffusion
problem is given by 
\begin{equation}
C(x,t)=\sum_{n=1}^\infty A_n \sin \frac {n\pi x}{L} 
\, e^{-\left(\frac{n\pi } L\right)^2 D t}
\end{equation}
with
$A_n = \frac 2 L \int_0^L C(x,0) \sin \frac {n\pi x}{L} dx$
and $n=1,2,\dots$ etc.
The absorbing boundary conditions are represented by $C(0,t)=C(L,t)=0$.

In this case the quiet time turns out to be
just the first-passage at $x=0$. 
If the concentration $C(x,t)$ is normalized to 1 in $t=0$,
the probability that the first-passage time %%(to any of the two boundaries) 
is larger than a value $t$ is 
\begin{equation}
S(t)=\int_0^L C(x,t) dx,
\end{equation}
which corresponds to the fraction of Brownian particles
that have not left the system,
and therefore, the probability density of the first-passage time
is 
\begin{equation}
f(t) = - \frac{dS(t)}{dt}=-\int_0^L \frac{\partial C(x,t)}{\partial t} dx =
J(L,t)-J(0,t)
\end{equation}
where $J(x,t)=-D \frac{\partial C}{\partial x}$ is the flow of particles.
This $f(t)$ is the probability density of the first passage time to any of the two boundaries.
The quiet-time distribution will be given by the out-flow of particles
at $x=0$ \cite{Redner}, which is then
\begin{equation}
f_L(t) = -J(0,t)=D \left.\frac{\partial C(x,t)}{\partial x} \right |_{x=0} =
\frac{\pi D}L \sum_{n=1}^\infty n A_n e^{- t/\tau_n}
\end{equation}
with the characteristic times $\tau_n = \frac {L^2}{\pi^2 D n^2}$.
The subindex $L$ in the density indicates the dependence with $L$,
not the position of the boundary.
Note that this distribution is not normalized, 
due to the out-flow of particles at the other boundary
(the normalization factor is explained in Ref. \cite{Redner},
but it is not relevant for our purposes).

The behavior of the quiet-time density is obtained immediately in the limit of
very large times, by the largest time scale $\tau_1$, so
\begin{equation}
f_L(t) \simeq \frac{\pi D  A_1} L \, e^{-t/\tau_1}
\end{equation}
for $t \gg \tau_1$, taking into account also the value of the constants
$A_n$, which we will see do not alter this limiting behavior.
For this we need a precise initial condition; in our case this is
$C(x,0)=\delta(x-x_0)$, with $\delta$ de Dirac's delta function
and $x_0$ the initial position, very close to the boundary at $x=0$
and fulfilling therefore $x_0 \ll L$. Substituting in $A_n$,
\begin{equation}
A_n=
 \frac 2 L  \sin \frac {n\pi x_0}{L},
\end{equation}
from which it is immediate that $n A_n e^{-t n^2/\tau_1}$
for $n=1$ is greater than any other term if $t \gg \tau_1$.
The fact that, in this limit, the term with $n=1$ is greater than the sum of the rest is made clear below.

Coming back to the quiet-time density for all times, 
we get, substituting the expression for $A_n$ given by the initial
condition,
\begin{equation}
f_L(t)=
\frac{2\pi D}{L^2} \sum_{n=1}^\infty n \sin \frac {n\pi x_0} L e^{-n^2 t/\tau_1}
\simeq
\end{equation}
\begin{equation}
\simeq \frac 2{\pi \tau_1} \int_1^\infty
dn n \sin (n\pi \ell) \,e^{-n^2 t/\tau_1},
\end{equation}
where we have approximated the sum by an integral
and have introduced the rescaled distance of the initial position to the threshold, 
$\ell = x_0/L$, which verifies $\ell \ll 1$.

As functions of $n$, the scale of the sinus is given by $1/\ell$
and the scale of the Gaussian factor by $\sqrt{\tau_1/t}$.
Then, at intermediate times, large enough such that 
 $\sqrt {\tau_1 /t} \ll 1 /\ell$ but small enough to keep 
 $t \ll \tau_1$ the variation of the Gaussian factor is much faster
than that of the sinus and we can approximate the later
as $\sin (n\pi \ell) \simeq n \pi \ell$, 
and the resulting integral can be straightforwardly solved,
yielding,
\begin{equation}
f_L(t)\simeq
 \frac {2\ell}{\tau_1} \int_1^\infty
dn n^2 e^{-n^2 t/\tau_1} =
\end{equation}
\begin{equation}
= \frac{\ell} {\tau_1} \left(\frac {\tau_1}{t}\right)^{3/2} \int_{t/\tau_1 \simeq 0}^\infty du \sqrt{u} \,e^{-u} = \frac {\sqrt{\pi} \ell }{2\tau_1 } \left(\frac {\tau_1}{t}\right)^{3/2}
\end{equation}
with the change  $n= \sqrt{\tau_1 u / t} $ and recognizing the gamma function of 3/2, which is $\Gamma(3/2) = \sqrt{\pi}/2$.
We repeat that this is valid in the range $\ell^2 \tau_1 \ll t \ll \tau_1$.

In the same way, we can go back to the case $t\gg \tau_1$
and compare the first term, $n=1$, with the sum of the rest, 
to see that
\begin{equation}
\frac {2\ell}{\tau_1}e^{-t/\tau_1} 
\gg  \frac{\ell} {\tau_1} \left(\frac {\tau_1}{t}\right)^{3/2}\Gamma(3/2,2^2 t/\tau_1) \sim \frac {2\ell} t e^{-4 t/\tau_1},
\end{equation}
using the incomplete gamma function, $\Gamma(\gamma,z)=\int_z^\infty y^{\gamma-1} e^{-y}dy$
and its asymptotic behavior $\Gamma(\gamma,z) \sim z^{\gamma-1} e^{-z}$
when $z\rightarrow \infty$.
Finally, at very short times
$\sqrt{t/\tau_1} \ll \ell$, 
the variation of the sinus is very fast compared with the Gaussian factor
and then the integral vanishes, approximately, so
\begin{equation}
f_L(t) \simeq \frac 1 {\tau_1} \int_1^\infty dn \sin (n\pi \ell) \cdot \mbox{ constant} \simeq 0.
\end{equation}

Summarizing, 
\begin{equation}
f_L(t) \simeq
\left\{\begin{array} {lll}
0 & \mbox{ for } & t \ll \ell^2 \tau_1 = \frac{x_0^2}{\pi^2 D}\\
&&\\
\frac {\sqrt{\pi} \ell }{2\tau_1 } \left(\frac {\tau_1}t\right)^{3/2}
& \mbox{ for } & \frac{x_0^2}{\pi^2 D} \ll t \ll \tau_1 =\frac{L^2} {\pi^2 D}\\
&&\\
\frac {2 \ell} {\tau_1} e^{-t/\tau_1}& \mbox{ for } & t \gg \tau_1\\
\end{array}\right. 
\label{supereq}
\end{equation}
where we can identify a simple scaling form. Nevertheless, 
as $\ell = x_0/L =x_0/ \sqrt{\pi^2 D \tau_1}=\sqrt{m/\tau_1}$,
with $m=x_0^2 /(\pi^2 D)$ the minimum waiting time
(below it, the density is zero),
the scaling law turns out to be
\begin{equation}
f_L(t) \simeq \sqrt{\frac {m} {\tau_1^{3}}} G\left(\frac t {\tau_1}\right),
\end{equation}
for $t\gg m$, with $m$ fixed,
or in terms of the threshold value $L$, 
\begin{equation}
f_L(t) \simeq \frac {\sqrt{D^{3}m}} {L^3} G\left(\frac {D t} {L^2}\right).
\end{equation}
where the scaling function $G$ absorbs missing multiplicative constants.
The scaling function in Eq. (\ref{supereq}) can be further approximated to give,
in this particular case,
\begin{equation}
f_L(t)\simeq \frac{\sqrt{\pi}/4 + (t/\tau_1)^{3/2}}{(t/\tau_1)^{3/2}}
%\left( \frac{2\ell}{\tau_1}\right) 
%\left( \frac{2\sqrt{m}}{\tau_1^{3/2}}\right)  e^{-t/\tau_1},
\left( {2\sqrt{\frac m{\tau_1^3}}} %{\tau_1^{3/2}}
\right)  e^{-t/\tau_1},
\end{equation}
for $t >  m$. This function has the right asymptotic behavior for all temporal
ranges, although other ``fits'' are possible.
%where $m=x_0^2 /(\pi^2 D)$ is the minimum waiting time
%(below it, the density is zero).
This simple model justifies the scaling ansatz
for the quiet time distributions.

%\bibliographystyle{unsrt}
%%
%\bibliography{../words_ramon/p1_lemmas/biblio}
%%%%%%%%%%%%%%%%%%%%%%%%%%%%%%%%%%%%%%%%%%%%%%%%%%%%%%%%%%%%%%%%%%%%%%%%%%%%%%%%%%%%%%%
%%\bibliographystyle{unsrt}
%%%\section*{References}
%%\bibliography{biblio.bib}%
%%\bibliography{bib_28122012.bib}

\end{document}